\documentclass[twocolumn,showpacs,preprintnumbers,amsmath,amssymb]{revtex4}
\usepackage{tabularx,graphicx}\begin{document}
\newcommand{\beq}{\begin{equation}}
\newcommand{\eeq}{\end{equation}}
\newcommand{\beqn}{\begin{eqnarray}}
\newcommand{\eeqn}{\end{eqnarray}}
\newcommand{\bmath}{\begin{subequations}}
\newcommand{\emath}{\end{subequations}}
\newcommand{\bra}[1]{\langle #1|}
\newcommand{\ket}[1]{|#1\rangle}

\title{Proposed experimental test of an alternative electrodynamic theory  of superconductors}
\author{J. E. Hirsch\footnote{Tel.: +1 858 534 3931, $email$: jhirsch@ucsd.edu} }
\address{Department of Physics, University of California, San Diego,
La Jolla, CA 92093-0319}

\begin{abstract} 
An alternative form of London's electrodynamic theory of superconductors predicts that the electrostatic screening length is the same as the
magnetic penetration depth. We argue that experiments performed  to date do not rule out this alternative formulation and propose an experiment
 to test it. Experimental evidence in its favor would have fundamental implications for the understanding of superconductivity.\end{abstract}
\pacs{}
\maketitle 
 It is not generally recognized that the conventional London electrodynamic description of superconductors \cite{londonbook} involves $two$ $independent$ $assumptions$,
 and that an alternative plausible formulation exists that is consistent with the Meissner effect but unlike the conventional formulation allows for the presence of electric fields in the
 interior of superconductors  \cite{london1,chargeexp}. Here we argue that this alternative formulation has not been subject to experimental test, discuss why this an important question
 to settle, and propose an experiment to do so.
 
 The conventional derivation of London's electrodynamic equations for superconductors starts from the ``acceleration equation''
 \beq
 \frac{\partial\vec{v}_s}{\partial t}=\frac{e}{m_e}\vec{E}
 \eeq
with $\vec{v}_s$ the superfluid velocity, $\vec{E}$ the electric field and $e$ and $m_e$ the superfluid carriers' charge and mass.
The electric current $\vec{j}_s=en_s\vec{v}_s$, with $n_s$ the density of superfluid carriers, then obeys
\beq
 \frac{\partial\vec{j}_s}{\partial t}=\frac{n_s e^2}{m_e}\vec{E}=\frac{c^2}{4\pi \lambda_L^2}\vec{E}
\eeq
with $\lambda_L=(m_ec^2/4\pi n_s e^2)^{1/2}$ the London penetration depth. Taking the curl on both sides, using Faraday's law, integrating in time and setting the
integration constant equal to zero yields the  London equation
\beq
\vec{\nabla}\times \vec{j}_s=-\frac{c}{4\pi \lambda_L^2}\vec{B}
\eeq
which, when combined with Maxwell's equation $\vec{\nabla}\times \vec{B}=(4\pi/c)\vec{j}_s$ yields
\beq
\nabla^2\vec{B}=\frac{1}{\lambda_L^2}\vec{B}
\eeq
and hence predicts that magnetic fields can only penetrate a superconductor
up to a distance $\lambda_L$ from the surface.

Integration of Eq. (3) yields
\beq
\vec{j}_s=-\frac{c}{4\pi \lambda_L^2}\vec{A}
\eeq
where  $\vec{A}$ is the magnetic vector potential. Taking the time derivative on both sides of Eq. (5) and using Faraday's law yields
\beq
 \frac{\partial\vec{j}_s}{\partial t}=\frac{c^2}{4\pi \lambda_L^2}(\vec{E}+\vec{\nabla} \phi)
 \eeq
 where $\phi$ is the electric  potential.  
 Eq. (6) differs from Eq. (2) in that it  allows for the presence of an electrostatic field in the interior of a superconductor, which, since
 $\vec{E}=-\vec{\nabla}\phi$, will not give rise to an infinite current as Eq. (2) predicts. London and London \cite{london1} pointed out that Eq. (3) has a greater degree of generality
 than Eq. (2) does, in other words that Eq. (2) can be derived from Eq. (3) only under the $additional$ $independent$ $assumption$ that $\vec{\nabla} \phi=0$
 in the interior of the superconductor, or equivalently that no electrostatic fields exist inside the superconductor. Note also that 
 Eq. (1), from which Eq. (2) was derived,
 does not follow from Newton's equation, rather Newton's equation yields Eq. (1) with the $total$ time derivative rather than the partial time derivative on
 the left side.  As a consequence, Eq. (6) $is$ compatible with Newton's equation \cite{chargeexp}.
 
The conventional formulation of London electrodynamics \cite{londonbook}  $assumes$ $\vec{\nabla}\cdot \vec{A}=0$ in Eq. (5), which implies that no electric field nor charges can exist inside superconductors. The alternative formulation assumes that $\vec{A}$ in Eq. (5) obeys the Lorenz gauge and leads to the following equations for the charge density and electrostatic field in the interior
of superconductors:
\bmath
\beq
\nabla^2(\rho-\rho_0)=\frac{1}{\lambda_L^2}(\rho-\rho_0)
\eeq
\beq
\nabla^2(\vec{E}-\vec{E}_0)=\frac{1}{\lambda_L^2}(\vec{E}-\vec{E}_0)
\eeq
\emath
with either $\rho_0=0, \vec{E}_0=0$ \cite{london1,fabrizio,bertrand,astro,badia,proca}, or 
$\rho_0$ a positive constant, with $\vec{\nabla}  \cdot \vec{E}_0=4\pi \rho_0$ \cite{chargeexp}. $\rho_0>0$ implies that the charge distribution in superconductors is
macroscopically inhomogeneus, with excess negative charge near the surface and a radial electric field in the interior \cite{chargeexp}.

Eq. (7) with either $\rho_0=0$ or $\rho_0\neq 0$ implies that the screening length for applied electrostatic fields
in superconductors is $\lambda_L$, typically several hundreds $\AA$, much longer than the Thomas Fermi screening length in normal metals, typically of order $\AA$.
H. London attempted to test the validity of Eq. (7)  experimentally \cite{londonexp} by looking for changes in the capacitance of a capacitor where the mercury electrodes changed from normal to
superconducting as the temperature is  lowered. He hypothesized that if the electric field penetrates a distance $\delta\sim \lambda_L$ into each electrode, the effective distance between electrodes would be
increased by $\sim 2\lambda_L$, leading  to a measurable decrease in the capacitance. He detected no change, and based on this result the London brothers concluded
\cite{londonexp,london2}  that the electric field does not penetrate a superconductor, hence that conventional London electrodynamics,
with $\vec{\nabla} \cdot \vec{A}=0$, describes superconductors in nature. 

In this paper we argue that H. London's test could not have detected whether Eq. (7) is valid. Furthermore we argue than no subsequent experiment has tested Eq. (7). Finally we propose
an experiment that can rule out or confirm Eq. (7).

Consider a superconducting electrode in a capacitor subject to a uniform electric field $E$ normal to its surface
as shown in Figure 1. We argue that the negatively charged
superfluid as a whole will rigidly shift with respect to the positive ionic background creating a surface charge density $\sigma$ that will  prevent the electric field from penetrating the interior. 
Using $E=4\pi \sigma$, $\sigma=en_s\delta$ we find
\beq
\delta  =\frac{E}{4\pi en_s} =\frac{eE}{m_ec^2}\lambda_L^2
\eeq
so e.g. for an applied electric field of $10^5 V/cm$ and a typical London penetration depth $\lambda_L= 500 \AA$ the displacement required to screen 
the electric field is a tiny $\delta=4.5\times 10^{-4} \AA$.
Therefore, the electric field will not penetrate the superconducting electrode and no change in the capacitance will be detected when the
electrode goes from the normal to the superconducting state. Thus, the null result of  H. London's experiment is explained independent of the validity or invalidity
of Eq. (7). Similarly, the null results of two recent experiments designed to test Eq. (7) \cite{bertrandexp,deheer} are explained by Fig. 1.

It has been argued \cite{ray} that experiments with single electron transistor devices \cite{set} (SET's) performed in recent years  \cite{tinkham,devoret,martinis} should have detected the unconventional
behavior predicted by Eq. (7) if it existed. A SET consists of a small metallic island connected to leads through 
small-capacitance tunnel junctions, and these experiments have been performed with superconducting $Al$ at temperatures well below the transition
temperature \cite{tinkham,devoret,martinis}. The charging energy of
the island is inversely proportional to the sum of the capacitances of the tunnel junctions involved, and would undergo an appreciable change if electric fields were to 
penetrate a London penetration depth when the system is cooled, and such changes have not been reported in the literature \cite{ray}.  However, we argue that the geometry
of these devices  \cite{set2} is such that the electric fields are uniform over distances much larger than the London penetration depth, hence
 it allows for a rigid shift of the superfluid to screen the electric fields as shown in Fig. 1, and consequently these experiments have nothing to
say about the validity or invalidity of Eq. (7).

   \begin{figure}
 \resizebox{8.5cm}{!}{\includegraphics[width=6cm]{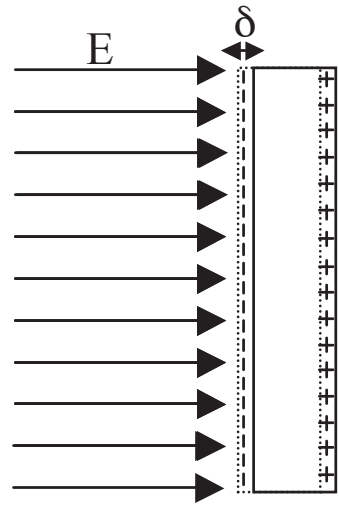}}
 \caption { Capacitor plate made of a superconducting material (solid rectangle). When a uniform electric field pointing towards the plate is applied, 
 the negative superfluid will shift rigidly a distance
 $\delta$ to the left to nullify the electric field in the interior. The dotted line denotes schematically the boundary of the superfluid.  }
 \label{figure1}
 \end{figure}
 
Similarly, it has been argued  \cite{doug} that experiments with superconducting microwave resonators performed in recent years \cite{stripline,kumar,barends} prove the invalidity of Eq. (7). 
The resonance frequency of these devices is inversely proportional to the square root of the capacitance of the system and should show different behavior at
low temperatures than what is seen experimentally if electric fields penetrate the superconducting components a distance $\lambda_L$ \cite{doug}. 
However, again we argue that because  for these devices the electric field applied to the   superconducting components is uniform over distances much
larger than the London penetration depth,  a rigid shift of the superfluid as shown
in Fig. 1 will prevent the electric field from penetrating the superconducting components, thus not testing the validity or invalidity of Eq. (7).

   \begin{figure}
 \resizebox{7.5cm}{!}{\includegraphics[width=6cm]{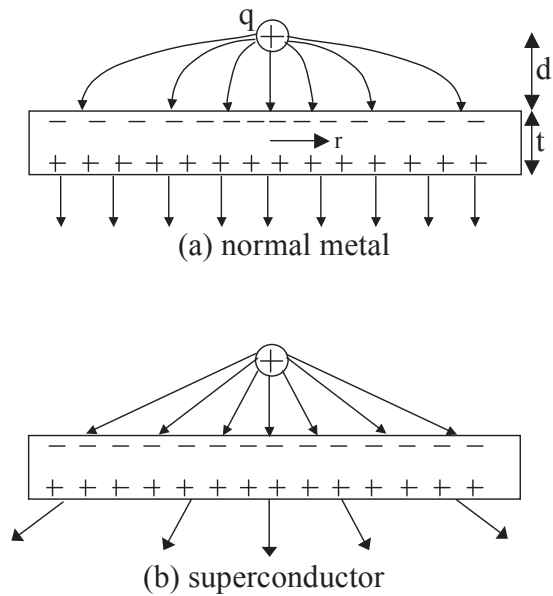}}
 \caption {Charge $q$ at distance $d$ from a (a) a normal metal and (b) a superconducting film. The lines with arrows are electric field lines. The electric field lines in the
 interior of the superconductor are not shown. The electric field is zero in the interior of the normal metal. }
 \label{figure1}
 \end{figure} 

To test the validity of Eq. (7) it is necessary to apply an electric field that varies over distances smaller than the
London penetration depth. Consider the situation depicted in Figure 2. A positive test charge $q$ is placed at a distance
$d$ above a metallic  film of thickness $t$. When the metal is in the normal state, a non-uniform
surface charge density is induced on the upper surface, given by
\beq
\sigma(r)=-\frac{d q}{2\pi (r^2+d^2)^{3/2}}
\eeq
where $r$ is the radial coordinate in the plane of the film, with $r=0$ the position of the test charge. On the bottom 
surface of the normal metal film a $uniform$ positive charge density is induced which produces an electric field pointing
downward $normal$ to the surface of the film. The electric field is zero in the interior of the film, and the electric potential is constant everywhere on the film.
The induced charge density on the upper surface, Eq. (9), changes over a radial distance of order $d$. We assume
that $d$ is larger than the Thomas-Fermi length which is typically of order $1\AA$.
When the metal is cooled into the superconducting state, if Eq. (7) is valid $and$ $d$ is  much smaller
than $\lambda_L$ the surface charge density will be different than in the normal state, since it cannot vary over distances much
shorter than $\lambda_L$.

Consider in particular the simple situation where the horizontal linear dimension of the film is of order
$\lambda_L$, $d<<\lambda_L$ and also the thickness of the film $t$ is much smaller than $\lambda_L$. 
The induced charge density on the top surface will be essentially uniform
if Eq. (7) is valid since it cannot change over distances smaller than $\lambda_L$. The induced positive charge density on the bottom surface will also be uniform, just as in the normal metal. Thus the induced
charge density will only partially screen the electric field in the interior of the superconducting film, 
and below the bottom surface the electric field lines originating from the point charge will emerge as if the
superconducting film wasn't there. The most important qualitative change is that there will now be
an electric field component  $tangential$ to the film below the bottom surface
and as a consequence, the electric potential below the bottom surface will change in the direction parallel to the film surface, in contrast to the normal film.

 \begin{figure}
 \resizebox{8.5cm}{!}{\includegraphics[width=6cm]{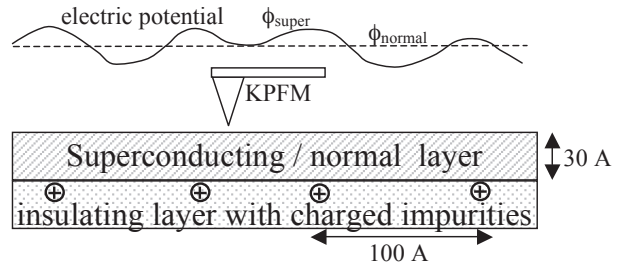}}
 \caption {A  Kelvin probe force microscope (KPFM) will detect variations in the electric potential $\phi_{super}$ above a thin superconducting film originating in
 charged impurities below the film if the superconductor
 does not screen over distances smaller than $\lambda_L$. Instead, if the superconductor reverts to the normal state or if the superconductor
 is described by the conventional theory the electric potential from
 the impurities will be screened and the KPFM will measure a constant potential $\phi_{normal}$ (dashed line).}
 \label{figure1}
 \end{figure}

As a concrete experimental realization, consider the setup in figure 3. A superconducting or normal metal layer rests on top of an insulating layer. The superconducting material, e.g. $Pb$ or $Nb$, has London penetration depth $\lambda_L\sim 400\AA$ and the thickness of the layer
is $\sim 30\AA$, much smaller than the London penetration depth but large enough to have a sharp superconducting transition with $T_c$ close to the bulk value. 
The insulating layer contains charged impurities at random distances from each other, of order e.g.
$100\AA$, also much smaller than $\lambda_L$. A Kelvin probe force microscope \cite{kpfm} (KPFM) is used to probe the electric potential right above the superconducting layer, in
non-contact mode, i.e. the tip does not touch the surface of the layer. Alternatively, a very thin insulating layer could be deposited on the superconducting layer to
prevent electric contact with the KPFM tip. The KPFM will image the electric potential at the surface.  If the superconducting material is in the normal state
the electric field created by the charged impurities in the insulating layer will be uniform and normal to the metallic layer (Fig. 2a) and the KPFM will show no potential
variations along the surface except those associated with atomic potential variations over short length scales ($\sim\AA$). The same will be true when the metal becomes superconducting if the superconductor obeys the
conventional theory which predicts that it screens electric fields over the same distances as the normal metal, i.e. the Thomas-Fermi length of order $1\AA$.
Instead, if Eq. (7) is valid, the metallic layer in the superconducting state will not be able to screen the tangential components of the electric field created by the 
charged impurities, and the KPFM will detect slowly varying electric potential variations extending over 10's or 100's of Angstrom, as shown schematically in Fig. 3.
For example, for two charged impurities with charge $|e|$   at distance $100 \AA$ from each other, the electric potential measured by the KPFM at distance
$30 \AA$ above the impurities will drop by $0.1V$ as the KPFM tip is moved from being on top of one impurity to the midpoint between the two impurities.

However, it is important to take into account that Eq. (7) describes only the superfluid behavior. At finite temperatures there will also be normal quasiparticles
which will screen over the much shorter Thomas Fermi length. The effective screening length at finite temperatures
is given by \cite{rigidity}
\beq
\frac{1}{\lambda_{eff}(T)^2}=\frac{1}{\lambda_L^2} \frac{n_s(T)}{n}+\frac{1}{\lambda_{TF}^2}\frac{n_n(T)}{n}
\eeq
 where $\lambda_L$ and $\lambda_{TF}$ are the (zero temperature) London penetration depth and Thomas-Fermi screening length, and $n_s(T)$, $n_n(T)$ are the
 superfluid and normal fluid densities, and $n$ is the total carrier density. 
 In order for the second term in Eq. (10) to be smaller than the first, the condition
 \beq
 \frac{n_n(T)}{n}<\frac{\lambda_{TF}^2}{\lambda_{L}^2+\lambda_{TF}^2}\sim (\frac{\lambda_{TF}}{\lambda_L})^2
 \eeq
needs to be satisfied. The normal fluid density at low temperatures is given by
\beq
\frac{n_n(T)}{n}=\sqrt{2\pi} (\frac{\Delta_0}{kT})^{1/2}e^{-\Delta_0/kT}
\eeq
with $\Delta_0$ the energy gap. Assuming the BCS relation $2\Delta_0/kT_c=3.53$ and $\lambda_{TF}=1\AA$, $\lambda_L=400\AA$ as appropriate for
$Nb$ or $Pb$ yields the condition
\beq
T/T_c<0.124
\eeq
as the temperature range where the effective electric screening length will be larger than $\lambda_L/\sqrt{2}$.

The alternative Eqs. (7) can be understood as originating in a generalized rigidity of the superconducting wavefunction \cite{london1,chargeexp,rigidity}. In the conventional
understanding, rigidity means that the wavefunction of the superconductor  does not change under changes in the magnetic vector potential and leads to Eq. (5). 
In a relativistically covariant formulation it is natural to assume that   the wavefunction is also
rigid  under  changes in the electric potential, leading to \cite{london1,chargeexp}
\beq
\rho-\rho_0=-\frac{1}{4\pi \lambda_L^2}(\phi-\phi_0)
\eeq
 with $\nabla^2\phi_0=\rho_0$, from which Eq. (7) follows. This argument (with $\rho_0=\phi_0=0$) was already proposed in London's original work \cite{london1}.
 It was extraordinarily prescient that the London brothers invoked the Klein-Gordon equation, applicable to relativistic spin $0$ $bosons$, to justify Eq. (14) \cite{london1}, at a time when the
 bosonic nature of the superconducting charge carriers was $not$ understood. In their paper the London brothers talk about the wavefunction $\Psi$ described
 by the Klein-Gordon equation as the wavefunction ``of a single electron''.

 If experiments such as the one depicted in Fig. 3 detect the variation of electric potential along the plane direction predicted by Eq. (7), the next step will be to determine whether
 $\rho_0=0$ or $\rho_0\neq 0$. Experimentally this could be done by detecting spontaneous electric fields in the vicinity of non-spherical superconducting microparticles that
are predicted to arise   from the non-homogeneous charge distribution that results  if $\rho_0\neq 0$ \cite{ellip}.  However from a theoretical point of view we argue that $\rho_0= 0$ is
untenable for the following reason: if $\rho_0=0$, the reference frame where the superconducting body is at rest is not distinguished from any other
 inertial reference frame within this relativistically covariant theory \cite{london1}. This does not make physical sense,
 since the reference frame where the body is at rest is clearly special.   Instead, if $\rho_0\neq 0$ the inertial reference frame at rest with respect to the body is distinguished by the fact that it is the only
 reference frame where the spatial components of the four-current $J_\mu=(\vec{j}_s,ic\rho)$ are identically zero in the deep interior of the superconductor. Thus, $\rho_0\neq 0$ is a necessary condition for  the theory to represent physical reality.
 The fact that $\rho_0>0$ rather than $\rho_0<0$ follows from the fundamental charge asymmetry of matter,
 manifested in superconductors by the fact that rotating superconductors exhibit a magnetic field always $parallel$,
 never antiparallel, 
 to their angular velocity \cite{londonmoment}. The numerical value of $\rho_0$ is given in Ref. \cite{electrospin}.
 
 Within our theory, the extended rigidity of the superconducting wavefunction arises from the fact that superconducting carriers reside in overlapping
 orbits of radius $2\lambda_L$ \cite{electrospin}, which prevents short distance variations in the carrier density without disturbing the phase coherence \cite{rigidity}.
Experimental confirmation of the validity of Eq. (7) would call for a reexamination of the validity of BCS theory which predicts no electrostatic screening changes 
between the normal and the superconducting state \cite{bcs}. Experiments by W. A. de Heer and coworkers that detected large spontaneous electric dipole moments in $Nb$, $V$ and $Ta$ clusters 
at low temperatures \cite{deheer2} provide experimental support for the validity of Eq. (7).

  \acknowledgements{}
  The author is grateful to Ray Ashoori, Andrea Young, Doug Scalapino, John Martinis and Walter de Heer  for informative and stimulating discussions.

\end{document}